\documentstyle[prb,aps,multicol,psfig]{revtex}
\begin{document}
\draft
\tighten

\title{Kinetics of four-wave mixing for a 2D magneto-plasma
in strong magnetic fields}
\author{M. W. Wu and H. Haug}
\address{Institut f\"ur Theoretische Physik, J.W. Goethe Universit\"at
 Frankfurt, Robert-Mayer-Stra\ss e 8, D-60054 Frankfurt a. M.,
 Germany}
\date{\today ; E-mail: wu@mandala.th.physik.uni-frankfurt.de}
\maketitle
\begin{abstract}
We investigate the femtosecond kinetics of an
optically excited 2D magneto-plasma at intermediate and high
densities under a strong magnetic field perpendicular to the 
quantum well (QW).  We assume an additional weak
lateral confinement which lifts the degeneracy of the Landau levels partially.
We calculate the femtosecond dephasing and relaxation kinetics of the laser
pulse excited magneto-plasma due to bare Coulomb potential scattering,
because screening is under these conditions of minor importance.
In particular the time-resolved and time-integrated four-wave mixing
(FWM) signals are calculated
by taking into account three Landau subbands in both the valance and
the conduction band assuming an electron-hole symmetry.
The FWM signals exhibit quantum
beats mainly with twice the cyclotron frequency.
Contrary to general expectations, we find no pronounced slowing down
of the dephasing with increasing magnetic field. On the contrary, one obtains a
decreasing dephasing time because of the increase of the Coulomb matrix
elements and the number of states in a given Landau subband. In the
situation when the loss of scattering channels exceeds these increasing
effects, one gets a slight increase at the dephasing time.
However, details of the strongly modulated scattering kinetics depend
sensitively on the detuning, the plasma density, and the spectral pulse width
relative to the cyclotron frequency.  
\end{abstract}

\begin{multicols}{2}
\narrowtext

\section{Introduction}

Femtosecond pulse excitation in semiconductors induces
transient carrier populations, which can be studied through
ultrashort-pulse nonlinear optics to elucidate many-body effects,
such as time-dependent
Coulomb correlations. Numerous experimental and theoretical
studies have been devoted to this problem in the
magnetic-field-free case in the last ten years.\cite{proce,shah,haug}
Studies of femtosecond laser spectroscopy in the presence of
a strong magnetic field are relatively rare and are mainly focused on
magneto-excitons at low excitation densities
\cite{stafford,glut,siegner,wegener}.
For a strong resonant laser pulse which excites an electron-hole (e-h)
system with a density above the Mott ionization density,
the kinetics is dominated by Coulomb scattering in the correlated e-h plasma
or in the correlated magneto-plasma if a strong magnetic field is present.
The femtosecond quantum kinetics and the expected FWM signal of the
 non-equilibrium e-h plasma have been treated recently by Vu {\em et al.}
 \cite{vu} using bare Coulomb potential for times shorter than the build-up
 time of screening. 
While the density-dependence of the optical spectra of a
quasi-equilibrium 2D magneto-plasma have been calculated,\cite{bauer}
the kinetics of a magneto-plasma has been studied neither
experimentally nor theoretically. Moreover, the existing theory for
the kinetics with a magnetic field is also not yet
developed as far as in the field-free case. However, as experimental studies
of the relaxation and dephasing kinetics in QW's and superlattices
are in progress\cite{private}, we will provide a first calculation of the
femtosecond FWM signal of a resonantly excited magneto-plasma in a single
quantum well. We concentrate on the study of carrier-carrier scattering,
while phonon and disorder scattering is not considered.

A strong magnetic field perpendicular to the QW plane forces the
carriers on cyclotron orbits, thus causing an additional quantum confinement.
In a magnetic field of the order of more than 10T only a few Landau levels
both in the conduction and valence band have to be considered. The high
degeneracy of these Landau levels is partially lifted by spatial
inhomogeneities and by size effects. The numerical treatment of the
broadening due to disorder, e.g. interface fluctuations, would require a
stochastic averaging over many FWM 
signal calculations which is beyond present day numerical possibilities
for the complex carrier kinetics treated here.
Therefore, we lift the degeneracy partially by a weak parabolic
confinement potential. The Landau levels are broadened by this confinement
into Landau subbands with a band 
width of about 2 meV. This weak parabolic confinement allows to treat the
single-particle Schr\"odinger equation still exactly in the presence of the
magnetic field.\cite{bayer} These eigenfunctions of the 2D electron
provide a convenient expansion basis for the non-equilibrium many-body
problem.
A strong quantum confinement --- here caused by the QW and the strong
magnetic field --- is generally believed to slow down the relaxation kinetics
because of the reduction of the phase space for scattering
processes.\cite{weisbuch} Indeed, for low-density magneto-excitons
an increase in dephasing time has been observed.\cite{wegener}
However, this rule does not apply to
Coulomb scattering, as has been shown recently for the example of
exciton-exciton scattering in 
QW wires.\cite{braun} For decreasing wire width a reduction of the
dephasing time has been found and explained in terms of the increase of
the Coulombic interaction matrix elements which overcompensated
the reduction of phase space for the scattering processes. Therefore it is
not obvious, how the Coulomb intra- and inter-Landau-subband scattering
will influence the resulting dephasing time in a dense magneto-plasma
in detail.

We present a kinetic study for a femtosecond laser-pulse excited 2D
dense non-equilibrium magneto-plasma in a QW in the framework of the
semiconductor Bloch equations combined with Coulomb scattering rates (Sec. II).
We expand the density matrix of a two-band semiconductor in the eigenfunctions
of the 2D electron in the 
presence of the strong magnetic field and the weak parabolic confinement.
We formulate the scattering terms for the population distribution functions of
the various Landau subbands and for the optically induced polarization
components between the Landau-subbands in the conduction and valence band in
the form of non-Markovian quantum kinetic scattering integrals\cite{haug} 
and in the form of semiclassical Boltzmann-type scattering rates.
Pronounced quantum kinetic effects are expected for time scales shorter than
typical inverse frequencies. For the considered high magnetic fields, the
cyclotron frequencies are so large that quantum kinetic effects should be of
minor importance. Furthermore, we prefer the use of the simpler semiclassical
kinetics, because there exist still some conceptual difficulties for the
quantum kinetic description, connected with
the spectral electron Green functions in a strong magnetic field, as will be
discussed below. 
We calculate the time-resolved (TR)
and time-integrated (TI) four-wave mixing (FWM) signals for two 50 fs pulses
by taking into account up to three Landau subbands in both the valance
band and the conduction band. The carrier frequency of the two delayed pulses
is tuned slightly above 
the un-renormalized energy gap. We simplify the problem by assuming equal
effective electron and hole masses, as it can be approximately realized in
strained QW's. Naturally, unequally effective masses will
lead to more complicated 
quantum beat structures in the FWM signals and modify to some extend also the
resulting 
relaxation and dephasing rates. Thus our present study should be seen
only as a first idealized model calculation. The detailed numerical results
based on the semiconductor Bloch equations with Boltzmann-type scattering
kinetics are presented in Sec.\ III for intermediate and high plasma
densities. The FWM signal is calculated by an adiabatic projection technique
which is appropriate for thin samples where propagation effects are not
important. For many conditions pronounced quantum beat
structures with two times the cyclotron frequency are obtained both in the TR
and TI FWM signals. Furthermore, often 
relaxation oscillations are seen between the populations of two subbands,
before they relax to their stationary values. The dephasing times obtained
from the TI FWM signal are surprisingly short and strongly modulated as a
function of the magnetic field. The unexpectedly short resulting dephasing
times between 100 and 200 fs
depend not only on the field values, but also on the detuning, the pulse width
and the excited carrier densities. In spite of the remaining conceptual
difficulties we present 
also first quantum kinetic calculations and compare the FWM signals with those
obtained with Boltzmann-type kinetics in Sec.\ IV. 
A conclusion of our main results is given in Sec.\ V.
In an appendix we give a list of the analytically calculated Coulomb matrix
elements for the first three subbands.

\section{model and kinetic equations}
\subsection{Model and Hamiltonian}

We start our investigation of a quantum well in  the $x$-$y$ plane
which is further restricted by an additional weak lateral confinement
in $x$ direction which lifts the degeneracy of the Landau levels
partially and provides a weak inhomogeneous broadening.
The confinement is assumed to be given by a
harmonic oscillator potential for the
lower-lying states. This model has been used with a strong additional
confinement for the theory of the optical properties of thermal
magneto-plasmas in QW wires.\cite{bayer} A strong
magnetic field $B\ge$ 10T is applied perpendicular to the well. We assume
electron-hole symmetry, {\em i.e.} $m_e=m_h\equiv m$ with $m_e$ ($m_h$)
denoting the effective electron (hole)mass. Within the Landau gauge ${\bf
  A}=xB{\bf e}_y$ and 
effective mass approximation\cite{callaway} to the lowest QW
subband, the stationary 2D Schr\"odinger equation of a single electron
can be solved exactly ($\hbar=c=1$):
\begin{eqnarray}
&&\left [\frac{1}{2m}\left(-\frac{\partial^2}{\partial x^2}+(\frac{1}{i}
\frac{\partial}{\partial y}-exB)^2\right)\right.\nonumber\\
&&\mbox{}\hspace{1.cm}+\left.\frac{1}{2}m\Omega x^2-E_{cnk}
\right]\psi_{nk}(x,y)=0\;,
\end{eqnarray}
with the shifted Landau eigenfunction
\begin{equation}
\psi_{nk}(x,y)=\frac{e^{iky}}{\sqrt{L_y}}\phi_n(x-\delta x_k),
\end{equation}
where $\phi_n(x)$ is the eigenfunction of $n$-th Landau level ($n=0$, 1,
$\cdots$) and reads\cite{callaway}
\begin{equation}
\phi_n(x)=\left(\frac{\alpha}{\sqrt{\pi}2^n n!}\right)^{1/2}
H_n(\alpha x)e^{-\frac{1}{2}\alpha^2x^2}\;,
\end{equation}
with $H_n(x)$ standing for the $n$-th order Hermite polynomial and
$\alpha=\sqrt{m\Omega_x}$ is the inverse of the amplitude of the zero-point
fluctuations. $\Omega_x$ is the effective oscillation frequency
$\Omega_x=\sqrt{\Omega^2+\omega_c^2}$, where
$\omega_c=eB/m$ is the cyclotron frequency and
$\Omega$ is the frequency of the  additional confinement potential.
The shift $\delta x_k=-\omega_c k/(m\Omega_x^2)$ results from the balance between
the Lorentz force and the harmonic restoring force $e(k/m)B=
m\Omega_x^2\delta x_k$.

The single-particle energy spectrum of a conduction band
electron in $n$-th Landau
level and with momentum $k$  is given by
$E_{cnk}=E_g/2+\varepsilon_{nk}$ with
\begin{equation}
\label{ek}
\varepsilon_{nk}=\frac{\Omega^2}{\Omega_x^2}\frac{k^2}{2m}
+\Omega_x(n+\frac{1}{2})
\end{equation}
and $E_g$ standing for the energy gap.
The dependence of the energy on the momentum in $y$-direction results
from the additional confinement of the QW, which broadens each Landau level
into a small Landau subband.
The momenta are restricted to
\begin{equation}
\label{limitk}
|k|\leq L_xm\Omega_x^2/(2\omega_c)
\end{equation}
because the center of cyclotron should lie within the sample width
$L_x$.\cite{callaway}  It is interesting to notice that when
$\omega_c\gg \Omega$, the inhomogeneous broadening is of the same
for all magnetic fields as the increase of $k$ space in
Eq.\ (\ref{limitk}) is totally compensated by the increase of the
effective mass in Eq.\ (\ref{ek}).

With Coulomb interaction and with the interaction with a coherent
classical light field, the many-body Hamiltonian for the electrons in the
conduction and valence bands is in the basis of above given magnetic
eigenfunctions
\begin{eqnarray}
\label{hamilton}
H&=&\sum_{\nu n k} E_{\nu nk}c^\dagger_{ink}c_{ink}+\frac{1}{2}
\sum_{\nu_1\nu_2;nmij;kk^\prime q}V_{ni;jm}(q,k,k^\prime)\nonumber\\
&&\mbox{}\times c^\dagger_{\nu_1nk+q}
c^\dagger_{\nu_2ik^\prime-q}c_{\nu_2jk^\prime}c_{\nu_1mk}+H_I\;,
\end{eqnarray}
with $\nu=c$ and $v$ standing for the conduction band and the valence band
respectively. $E_{vnk}=-E_{cnk}$ is the energy spectrum of an electron
in the valence band.

The Coulomb interaction matrix elements $V_{ni;jm}(q,k,k^\prime)$
describes the scattering of an electron with momentum $k$ from $m$-th Landau
subband to $n$-th subband with momentum $k+q$
and an electron with momentum $k^\prime$ from $j$-th subband to $i$-th
with momentum $k^\prime-q$. Due to the assumed e-h
symmetry, these four Landau subbands can be
either in the conduction band or in the valence band. The matrix elements are
given in terms of the shifted magnetic eigenfunctions by the following integral
\begin{eqnarray}
&&V_{ni;jm}(q,k,k^\prime)=\sum_{q_x}\frac{2\pi e^2}{\epsilon_0
\sqrt{q^2+q_x^2}}\int dxdx^\prime
e^{-iq_x(x-x^\prime)} \nonumber\\
&&\mbox{}\hspace{0.2cm}\times
\phi_n^\ast (x-\delta x_k+\delta x_q)
\phi^\ast_i(x^\prime-\delta x_{k^\prime}
-\delta x_q)\phi_j (x^\prime-\delta x_{k^\prime})\nonumber\\
&&\mbox{}\hspace{0.2cm}\times
\phi_m (x-\delta x_k)\;,
\label{vo}
\end{eqnarray}
$\epsilon_0$ is the background dielectric
constant. After substituting $x\longrightarrow x+\delta x_k$ and
$x^\prime\longrightarrow x^\prime+\delta x_{k^\prime}$ in Eq.\ (\ref{vo}),
one gets
\begin{eqnarray}
&&V_{ni;jm}(q,k,k^\prime)=\sum_{q_x}\frac{2\pi e^2/\epsilon_0}{
\sqrt{q^2+q_x^2}}\int dxdx^\prime
\exp\{-iq_x[x-x^\prime \nonumber\\
&&\mbox{}\hspace{0.1cm}+\lambda(k-k^\prime)]\}
\phi_n^\ast (x+\delta x_q)
\phi^\ast_i(x^\prime-\delta x_q)\phi_j (x^\prime)\phi_m (x)\;,
\label{v}
\end{eqnarray}
with $\lambda=-m\omega_c/\alpha^2$.
After an approximation concerning the momentum dependence,
the integrals of Eq.\ (\ref{v}) may be carried out analytically.
In the appendix the corresponding approximation and the final expressions of
$V_{ni;jm}(q,k,k^\prime)$ in terms of zeroth- and first-order modified
Bessel functions are given.

$H_I$ in Eq.\ (\ref{hamilton}) denotes the dipole coupling with
the light field $E(t)$. In the assumed e-h symmetry it contains
only transitions between Landau subbands of the same order:
\begin{equation}
\label{hi}
H_I=-d\sum_{nk}E(t)(c^\dagger_{cnk}c_{vnk}+h.c.)\;.
\end{equation}
In this equation, $d$ denotes the optical-dipole matrix element.
The light field is further split into
$E(t)=E_0(t)e^{i\omega t}$ with $\omega$ being the central frequency
of the coherent pulse. $E_0(t)$ describes a Gaussian pulse
$E_0 e^{-t^2/\delta t^2}$ with $\delta t$ denoting the pulse width.

\subsection{Kinetic Equations}

Following the same scheme as for the magnetic-field-free
case in Ref.\ \onlinecite{haug}, appropriately modified,
one may build the semiconductor Bloch
equations for the QW in a strong perpendicular magnetic field $B$ in the basis
of the magnetic eigenfunctions as follows:
\begin{equation}
\label{eqs}
\dot \rho_{\nu ,n,\nu^\prime,n^\prime,k}=\left.\dot
\rho_{\nu ,n,\nu^\prime,n^\prime,k}\right |_{\mbox{coh}}
+\left.\dot \rho_{\nu ,n,\nu^\prime,n^\prime,k}\right |_{\mbox{scatt}}.
\end{equation}
Here $\rho_{\nu ,n,\nu^\prime,n^\prime,k}$ represents the
single-particle density matrix.
The diagonal elements describe the carrier distribution functions
 $\rho_{\nu ,n,\nu ,n,k}= f_{\nu nk}$ of the $n$-th
Landau subband and the wavevector $k$ as diagonal elements, and the
off-diagonal elements describe the interband polarization components, e.g.
$\rho_{c,n,v,n,k} = P_{nk}e^{-i\omega t}$.  For the
assumed e-h symmetry, $f_{enk}\equiv
f_{hnk}\equiv f_{nk}$ and the polarization has only components between
subbands of the same quantum number n in the conduction and valence band,
which simplifies the problem considerably.

The coherent part of the equation of motion for
the distribution function is in the rotating wave approximation given by
\begin{eqnarray}
\label{fcoh}
&&\left.\dot f_{nk}\right|_{\mbox{coh}}
=-2\mbox{Im}\{[dE_0(t)/2+\sum_{m,q}V_{nm;nm}(q,k,k)\nonumber\\
&&\mbox{}\hspace{1cm}\times P_{m k+q}(t)]P^\ast_{nk}(t)\}\;.
\end{eqnarray}
The first term describes the generation rate by the laser pulse. $d$ is
the optical dipole matrix element. For the assumed e-h symmetry transitions
between different subband quantum numbers are not allowed. The second term
describes the 
exchange interaction correction of the exciting laser by  the e-h attraction,
thus it can be seen
as a local field correction of the time-dependent bare Rabi frequency
$dE_0(t)$.  
The retarded quantum kinetic scattering rates for the considered bare Coulomb
potential scattering are given by\cite{haug,vu}
\begin{eqnarray}
\label{fscatt}
&&\left.\dot f_{nk}\right|_{\mbox{scatt}}
=-8\sum_{q k^\prime m i j}|V_{ni;jm}(q,k,k^\prime+q)|^2\nonumber\\
&&\mbox{}\hspace{0.5cm}\times\int_{-\infty}^t
dt^\prime \mbox{Re}\{e^{[-i(\varepsilon_{mk-q}-
\varepsilon_{nk}+\varepsilon_{jk^\prime+q}-\varepsilon_{ik^\prime})
-\Gamma](t-t^\prime)}\nonumber\\
&&\mbox{}\hspace{0.5cm}
\times[f_{nk}(t^\prime)f_{ik^\prime}(t^\prime)(1-f_{mk-q}(t^\prime))
(1-f_{jk^\prime+q}(t^\prime))\nonumber\\
&&\mbox{}\hspace{0.5cm}-f_{mk-q}(t^\prime)f_{jk^\prime+q}
(t^\prime)(1-f_{nk}(t^\prime))(1-f_{ik^\prime}(t^\prime))\nonumber\\
&&\mbox{}\hspace{0.5cm}
-P^\ast_{nk}(t^\prime)P_{mk-q}(t^\prime)(f_{ik^\prime}(t^\prime)
-f_{jk^\prime+q}(t^\prime))\nonumber\\
&&\mbox{}\hspace{0.5cm}-
P_{jk^\prime+q}(t^\prime)P^\ast_{ik^\prime}(t^\prime)(f_{nk}(t^\prime)
-f_{mk-q}(t^\prime))]\}\;.
\end{eqnarray}
In the derivation of this formula vertex corrections have been neglected. The
scattering self-energy is evaluated in time-dependent RPA. Successively, the
Coulomb potential is taken as a bare instantaneous potential and
the two-time-dependent particle propagators have been expressed in terms of
single-time density matrix elements and retarded (or advanced)
Green functions using the generalized Kadanoff-Baym ansatz.
For simplicity the retarded (and advanced) Green functions have been
approximated by a diagonal free-particle Wigner-Weiskopf form, i.e.
$G^r_{nk}(t,t^\prime)=G^{a\ast}_{nk}(t^\prime,t)
=-i\Theta(t-t^\prime)e^{[-i(E_g/2+\varepsilon_{nk})-\gamma](t-t^\prime)}$.
The effective damping
$\Gamma$ in Eq. (\ref{fscatt}) is the sum of the four imaginary parts of the
retarded electron self-energy which is assumed to be simply the
damping constant $\gamma$, so that $\Gamma\simeq 4\gamma$.
We will discuss later that this approximation for the imaginary part of the
retarded e-self-energy leads to some unphysical features in  the quantum
kinetics of the considered magneto-plasma. For further progress in the quantum
kinetics of 
a magneto-plasma, the calculations of the spectral functions have to be made
more self-consistent in order to include the retarded onset and the magnetic
field dependence of damping and the
band mixing by the coherent light pulses and the mean-field Coulomb
interactions. These refinements have been developed already for the simpler
case of scattering with optical phonons without magnetic
field.\cite{haug,banyai,hh} However, for the time being the
complexity of the Coulomb quantum kinetics in a magneto-plasma prevents us to
include these improvements in the present numerical evaluation.

The coherent time evolution of the interband polarization components are
\begin{eqnarray}
\label{pcoh}
&&\left.\dot P_{nk}\right|_{\mbox{coh}}=
-i\delta_n(k)P_{nk}(t)+i\Bigl[dE_0(t)/2\nonumber\\
&&\mbox{}+
\sum_{m,q}V_{nm;nm}(q,k,k)P_{mk+q}(t)\Bigr][1-2f_{nk}(t)]\;.
\end{eqnarray}
The first term gives the free evolution of the polarization components with
the detuning
\begin{equation}
\label{detuning}
\delta_n(k)=2\varepsilon_n(k)-\Delta_0
-2\sum_{m,q}V_{nm;nm}(q,k,k)f_{mq}(t)  
\end{equation}
with $ \Delta_0=\omega-E_g$\ .
$\Delta_0$ is the detuning of the center frequency of the light
pulses with respect to the unrenormalized band gap. The second term in
Eq. (\ref{pcoh} describes again the excitonic correlations in the
magneto-plasma, while the final factor describes the Pauli blocking. 

The dephasing of the polarization components are determined by the following
quantum kinetic scattering integral
\begin{eqnarray}
\label{pscatt}
&&\left.\dot P_{nk}\right|_{\mbox{scatt}}=
-4\sum_{qk^\prime,mij}|V_{ni;jm}(q,k,
k^\prime+q)|^2\nonumber\\
&&\mbox{}\times\int_{-\infty}^tdt^\prime\Bigl\{\Bigl[e^{[-i(
\varepsilon_{mk-q}+\varepsilon_{nk}+\varepsilon_{jk^\prime+q}-
\varepsilon_{ik^\prime}-\Delta_0)-\Gamma](t-t^\prime)}\nonumber\\
&&\mbox{}\times(P_{nk}(t^\prime)N_{mji}(k-q,k^\prime,q,t^\prime)-
P_{mk-q}(t^\prime)\nonumber\\
&&\mbox{}\hspace{1cm}\times N_{nji}(k,k^\prime,q,t^\prime))\Bigr]
-\Bigl[n\leftrightarrow m;\nonumber\\
&&\mbox{} i\leftrightarrow j;k
\leftrightarrow k-q;k^\prime\leftrightarrow k^\prime+q\Bigr]\Bigr\}
-\frac{P_{nk}(t)}{T_2}\;,
\end{eqnarray}
with the population factor
\begin{eqnarray}
&&N_{nji}(k,k^\prime,q,t^\prime)=(1-f_{nk}(t^\prime))(1-f_{jk^\prime+q}
(t^\prime))f_{ik^\prime}\nonumber\\
&&\mbox{}+f_{nk}(t^\prime)(1-f_{ik^\prime}(t^\prime))
f_{jk^\prime+q}(t^\prime)-P_{jk^\prime+q}(t^\prime)P^\ast_{ik^\prime}
(t^\prime)\;,
\end{eqnarray}
and the energy difference $\delta_n(k)=2\varepsilon_n(k)-\Delta_0
-2\sum_{m,q}V_{nm;nm}(q,k,k)f_{mq}(t)$.
$\Delta_0=\omega-E_g$ is the detuning with respect to the unrenormalized
band gap. Here $T_2$ is introduced phenomenologically
to describe additional slower scattering processes.
Eqs.\ (\ref{eqs})-(\ref{fscatt}), (\ref{pcoh}) and (\ref{pscatt}) are
the quantum kinetic Bloch equations for a magneto-plasma.

In the longtime limit the quantum kinetic scattering integrals be transformed
into Markovian Boltzmann-type scattering rates by pulling the slowly varying
distributions and polarization components
outside the scattering integrals at the upper limit $t$ and by
replacing the memory kernels $\int_{-\infty}^t dt^\prime
\exp\{[-i(\varepsilon_{mk-q}-
\varepsilon_{nk}+\varepsilon_{jk^\prime+q}-\varepsilon_{ik^\prime})
-\Gamma](t-t^\prime)\}$ in Eq.\ (\ref{fscatt}) by
$2\pi\delta(\varepsilon_{mk-q}-
\varepsilon_{nk}+\varepsilon_{jk^\prime+q}-\varepsilon_{ik^\prime})$
and $\int_{-\infty}^tdt^\prime \exp\{[-i(
\varepsilon_{mk-q}+\varepsilon_{nk}+\varepsilon_{jk^\prime+q}-
\varepsilon_{ik^\prime}-\Delta_0)-\Gamma](t-t^\prime)\}$ in
Eq.\ (\ref{pscatt}) by $2\pi\delta(\varepsilon_{mk-q}+\varepsilon_{nk}
+\varepsilon_{jk^\prime+q}-\varepsilon_{ik^\prime}-\Delta_0)$
in the limit of vanishing damping. In our calculations
we use for numerical convenience energy resonances with a small Gaussian
damping, which does not lead to such severe violations of the energy
conservation as the Lorentzian resonances do.   

It is noted that besides the e-h symmetry, we have further assumed that
Coulomb Auger scattering across the gap can be neglected.

\section{Numerical results based on Boltzmann kinetics}

We perform a numerical study of the Bloch equations in the Boltzmann
limit to study TR and TI FWM signals for high magnetic fields ($B>10$\ T)
for two degenerate Gaussian pulses of a width of 50 fs and a variable delay
time $\tau$. The pulses travel in the direction ${\bf k}_1$ and ${\bf k}_2$.  
We use an adiabatic projection technique in order to calculate
the polarization in the FWM direction with wavevector $2{\bf k}_2
-{\bf k}_1$ described in detail
in Ref.\ \onlinecite{banyai}.  This technique is suitable for optically
thin crystals, where the spatial dependence can be treated
adiabatically.\cite{koch} To do so, we replace the single-pulse
envelope function in Eq.\ (\ref{fcoh}) by two delayed pulses
$E_0(t)=E_0(t)+E_0(t-\tau)e^{i\varphi}$ with
the relative phase $\varphi=({\bf k}_2-{\bf k}_1)\cdot {\bf x}$
resulting from the different propagation directions. The projection technique
is used with respect to this phase. 
We further assume a resonant excitation with a fixed excess energy of
$\Delta_0= 26.4$\ meV. The intensity of each pulse will be given by
$\int_{-\infty}^{\infty} dE_0(t)dt=\chi\pi$, where $\chi$ denotes the fraction
of a $\pi$-pulse defined without local field corrections. 
Except $m_h$ which is assumed to be equal to $m_e$, all
the other material parameters are taken for bulk GaAs with an
exciton Rydberg of 4.2\ meV and a Bohr radius of 14\ nm. For the
sake of weak confinement,
the harmonic oscillator frequency $\Omega$ is assumed to be $0.1\omega_c$
for $B=7$\ T, corresponding to $\Omega=1.2$\ meV.
By fitting the energy $\Omega$ with the energy splitting
to the two lowest subbands of a rectangular square well
$3\pi^2/(2mL_x^2)$, we get the width in the additional
confinement direction to be $L_x=118$\ nm.\cite{bayer}
\begin{figure}[htb]
\psfig{figure=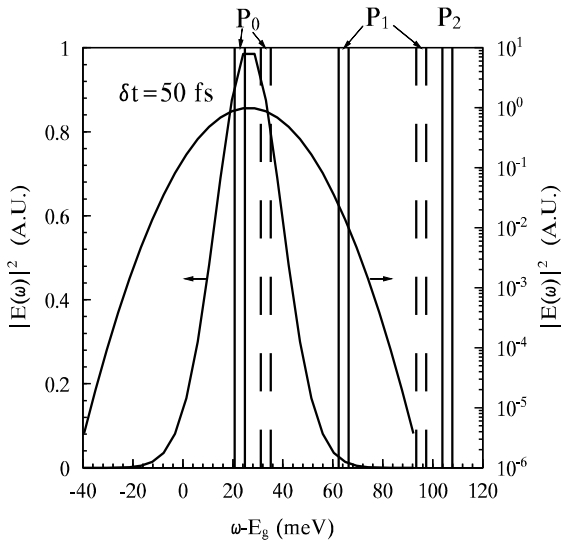,width=8.5cm,height=7.5cm,angle=0}
\caption{The pulse intensity spectrum $|E(\omega)|^2$ plotted in both
linear and log scale for a 50\ fs pulse, together with the unrenormalized
energybands for the optical transitions between Landau
subbands $n$(=0, 1, 2) plotted as solid lines for $B=12$\ T and as 
dashed lines for $B=18$\ T.}
\end{figure}

In Fig.\ 1 we plot the pulse spectrum together with the three transitions
bands for two magnetic fields of  $B=12$\ T
(solid lines) and 18\ T (dashed lines). The first transition band corresponding
to $P_0$ is due to transitions between the lowest Landau subband $n=0$
in the valance band to the
lowest subband $n=0$ in the conduction band. This transition band
starts an effective frequency $\Omega_x$ above the band gap at least in the
low-density limit.
Similarly the second transition band for the n=1  Landau subbands
in both valence and conduction bands requires
\begin{figure}[htb]
\psfig{figure=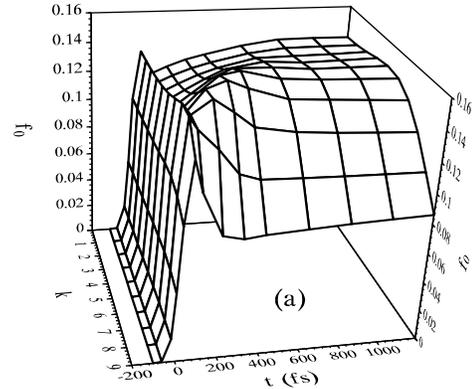,width=8.5cm,height=6.5cm,angle=0}
\psfig{figure=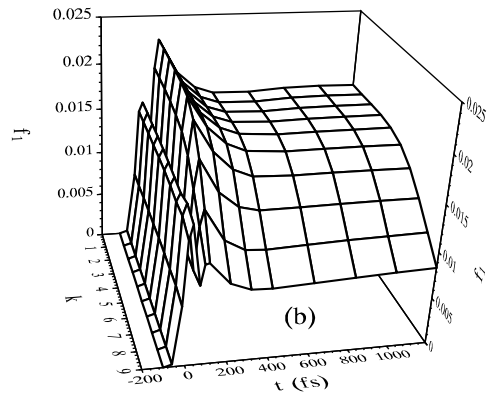,width=8.5cm,height=6.5cm,angle=0}
\psfig{figure=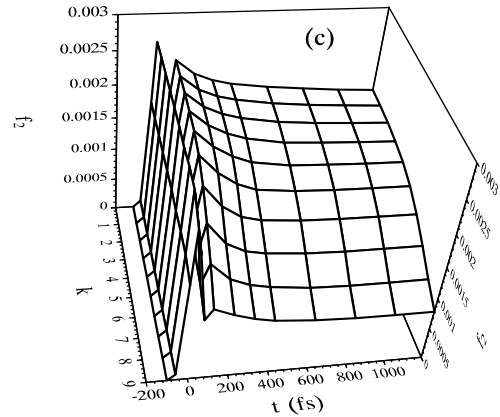,width=8.5cm,height=6.5cm,angle=0}
\caption{The distribution functions of three Landau subbands $f_0$, $f_1$
and $f_2$ are plotted as functions of $t$ and $k$ for a one-pulse
excitation with $B=12$\ T. The units of $k$ in the
figures are one tenth of the maximum allowed value of $k$ defined in
Eq.\ (\ref{limitk}).}
\end{figure}
\noindent the excess energy $3\Omega_x$.
The Hartree-Fock terms in Eq.\ (\ref{detuning})
may red-shift the energy of transition for the initial stage of excitation.
From Fig.\ 1 we see that for high magnetic fields, the pulse
mainly populates in the lowest Landau subband.
\begin{figure}[htb]
\psfig{figure=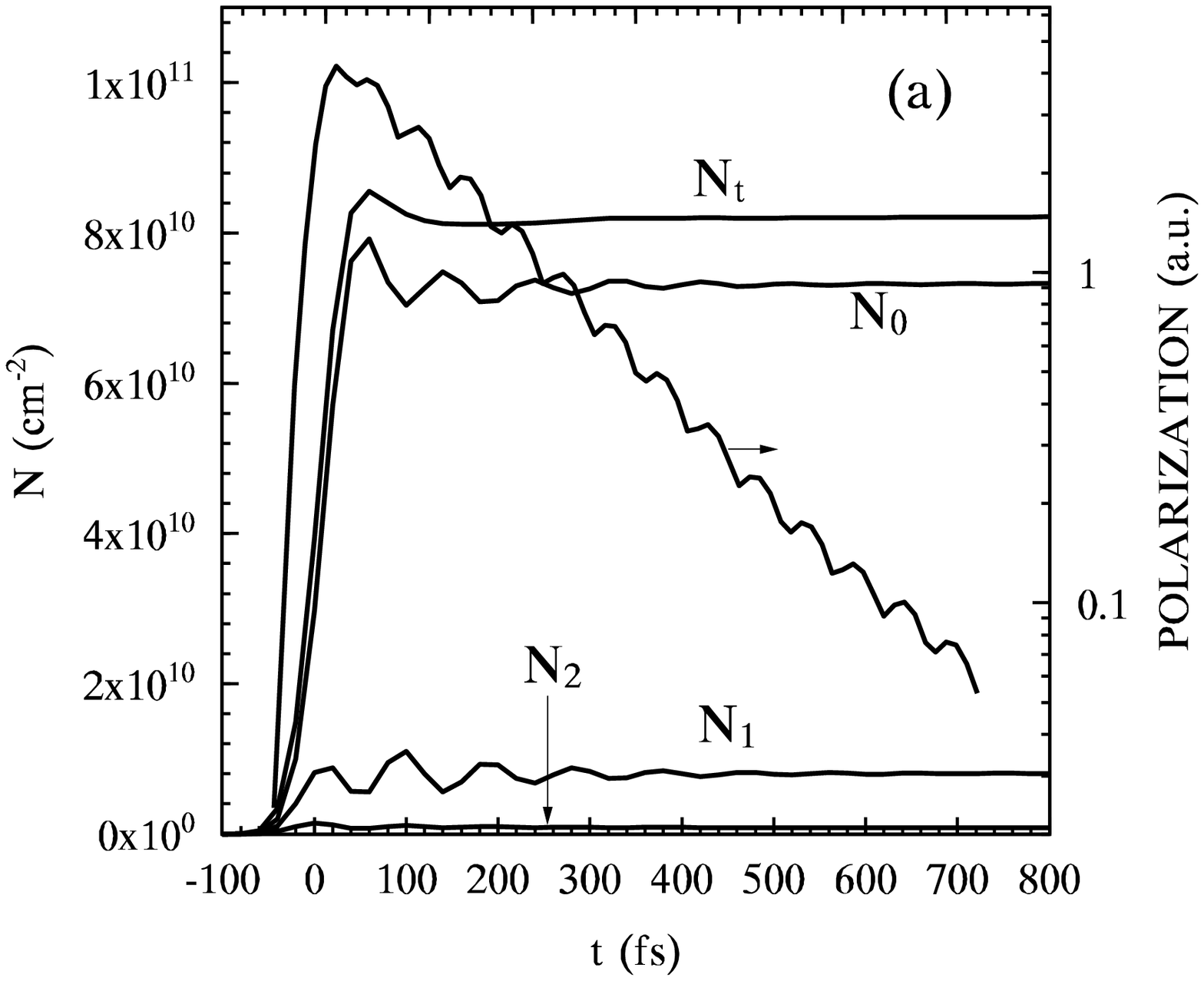,width=8.5cm,height=7.5cm,angle=0}
\psfig{figure=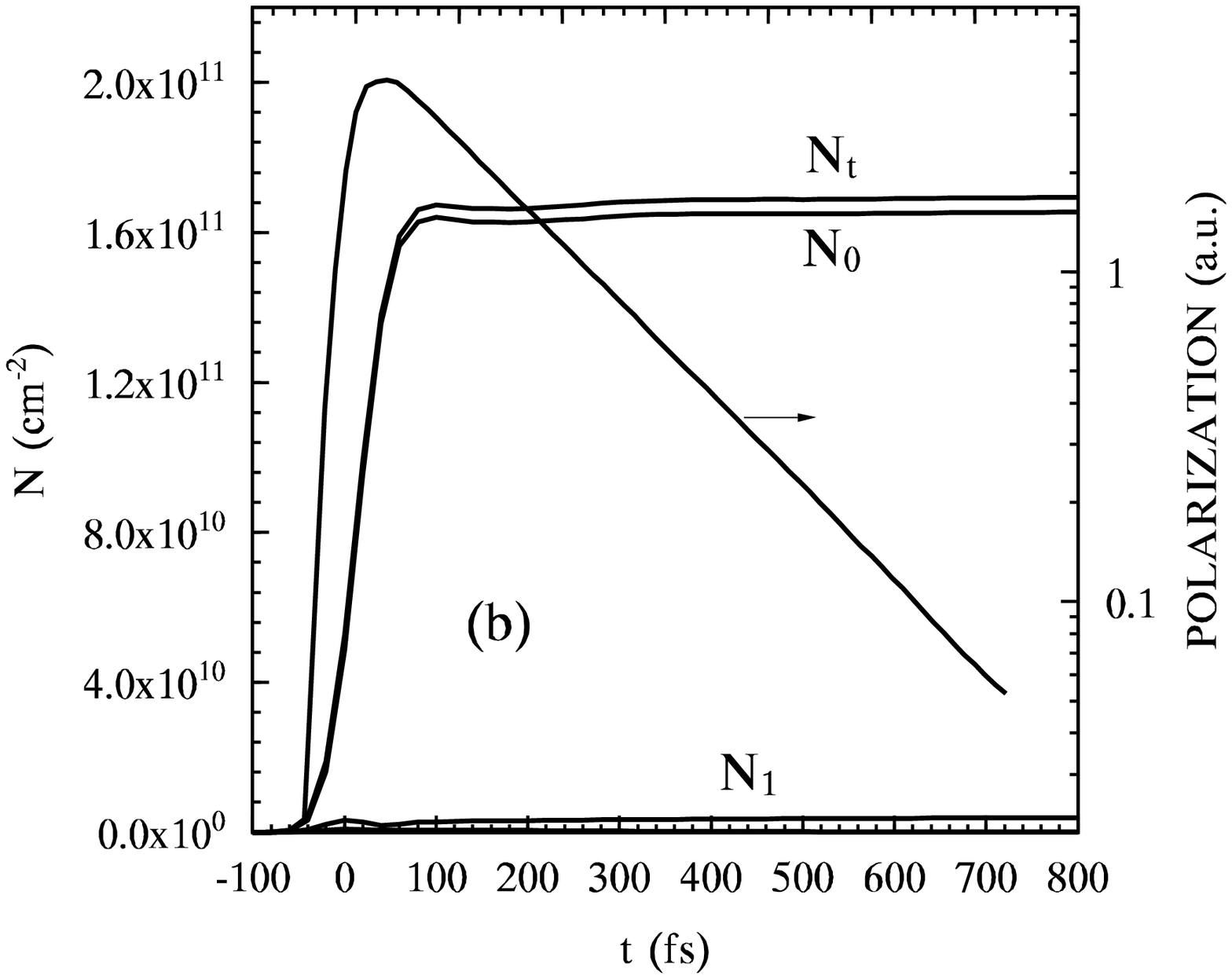,width=8.5cm,height=7.5cm,angle=0}
\caption{Electron densities of three Landau subbands $N_i$  and
the total density $N_t$ and the incoherently summed polarization
plotted against time $t$ for $B=12$\ T (Fig.\ 3(a)) and 18\ T
(Fig.\ 3(b)).}
\end{figure}

As mentioned before, we include in our calculations three Landau subbands
in the valence and conduction band.
It is important to include higher Landau subbands for the kinetics
after the excitation. In principle it is true even in situations where
the pulses mainly excite e's and h's only in the pair of the lowest Landau
n=0 subbands: A certain number of e's excited in the n=1 Landau subband can
only relax to the n=0 subband, if simultaneously the same number of e's
is excited from the n=1 to the n=2 subband if only three 
subbands are included. On the other hand, say if
one includes up to 5 Landau subbands,  then it is possible to only scatter
one electron from the second Landau subband to the fifth subband
\begin{figure}[htb]
\psfig{figure=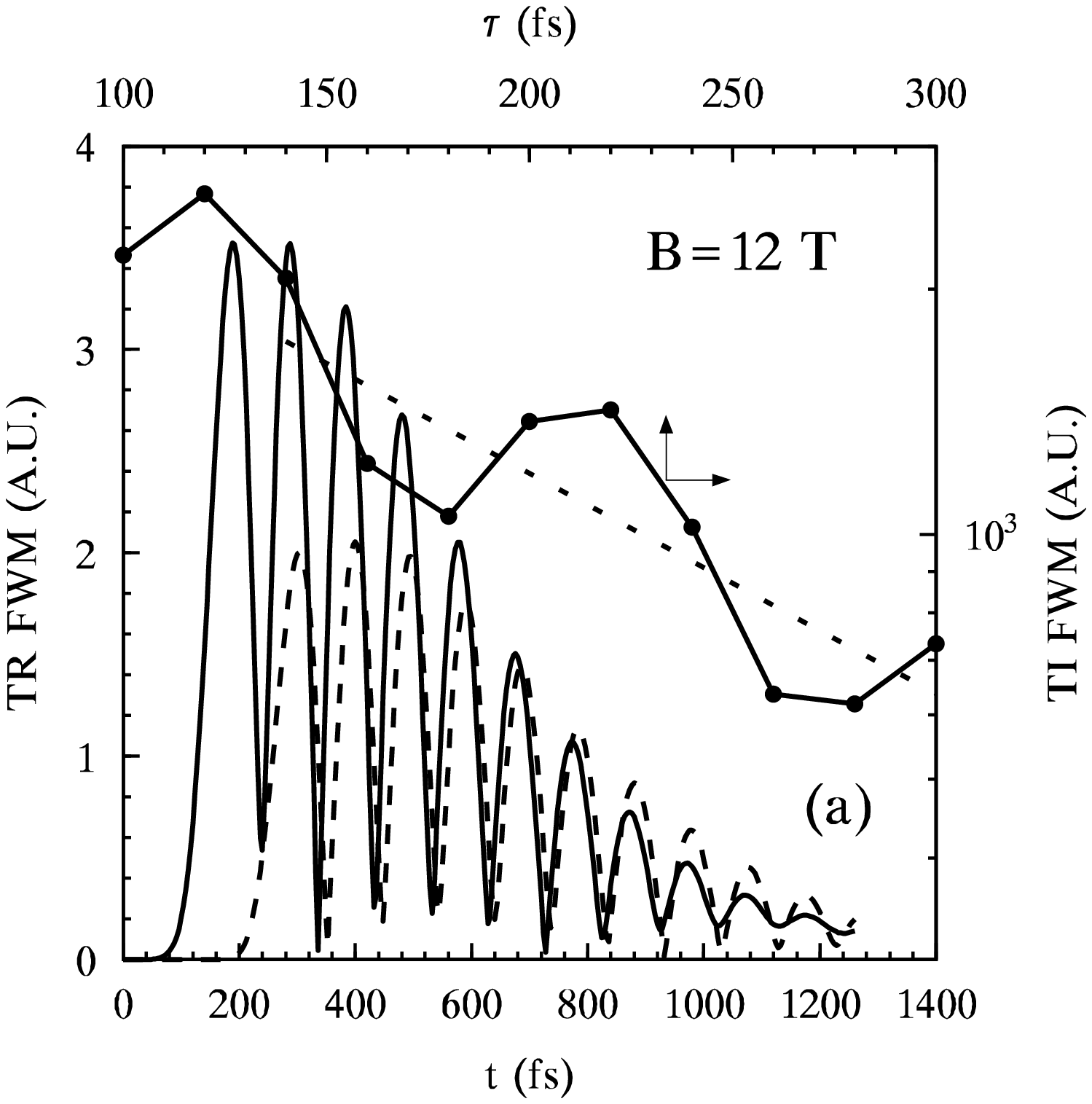,width=8.5cm,height=7.5cm,angle=0}
\psfig{figure=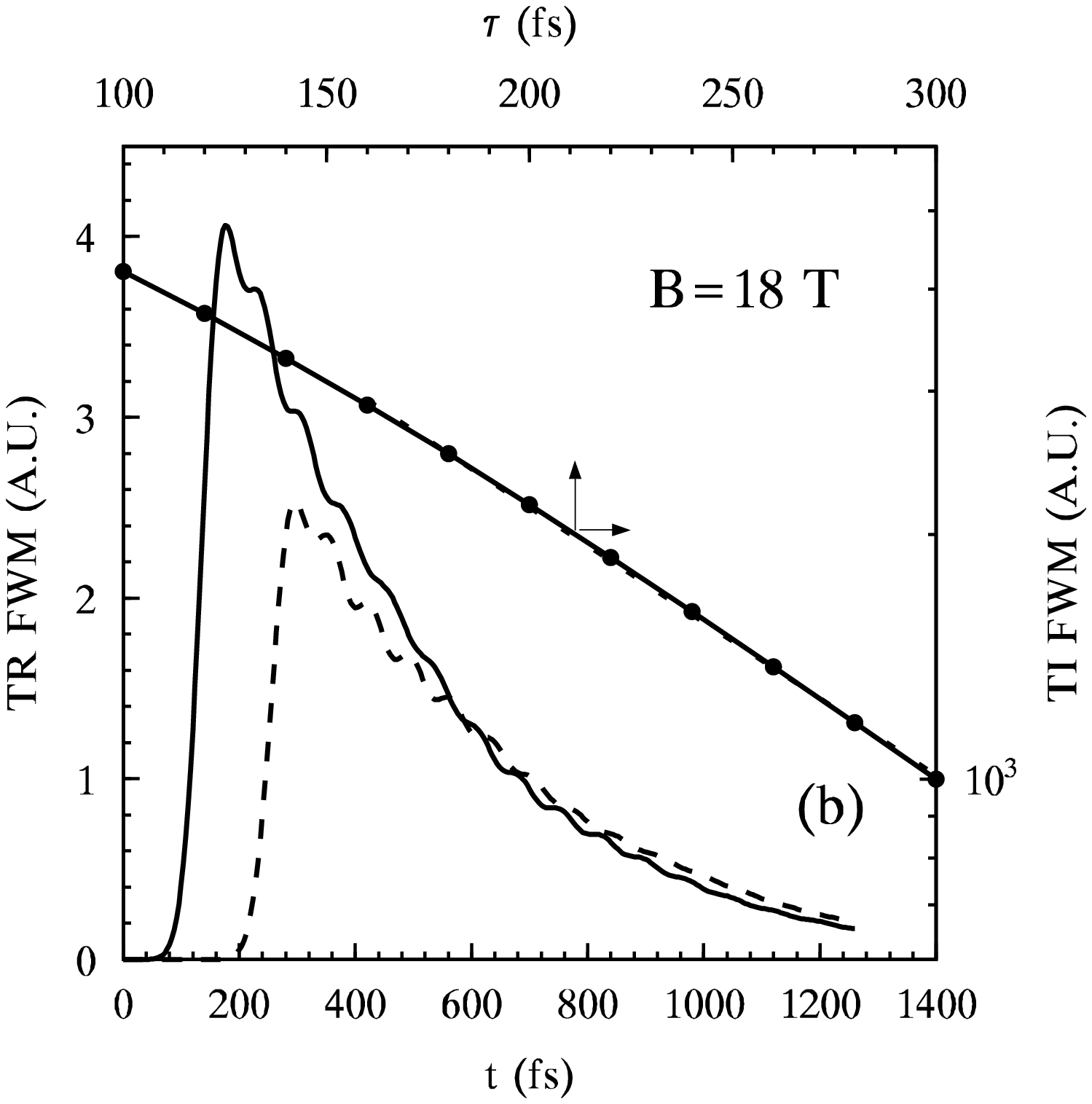,width=8.5cm,height=7.5cm,angle=0}
\caption{TR and TI FWM signals versus time $t$ and delay $\tau$,
respectively, for $B=12$\ T (Fig.\ 4(a)) and 18\ T (Fig.\ 4(b)).
For TR FWM, the solid curve is for $\tau=120$\ fs and dashed
curve for 240\ fs. The dotted line is
the exponential fit to the TI FWM signal. Note that the scale of
the delay time $\tau$ for TI FWM signals (top frame) is different from the
time scale of the TR FWM signals (bottom frame).}
\end{figure}
\noindent  and in the mean time to release three electrons from the second
subband to the lowest subband.
Therefore, a too small number of Landau subbands may 
change the Coulomb kinetics. For this reason, we restrict ourselves
to the high magnetic field regime in which for the chosen detuning
e's and h's are excited mainly in the lowest n=0 Landau subbands.
More Landau subbands are necessary in order to extend the kinetics
to lower magnetic fields or for larger detunings.
However, the expansion of number of matrix elements
$V_{ni;jm}$ increases as $N^4$ with $N$ being total number of Landau
subbands considered. With $N=3$ in our model, the number of form factors
is already 81, while it will be 256 with $N=4$.

\subsection{Intermediate-density case}

We first discuss the Boltzmann kinetics for an intermediate excitation
density. To do so, we choose a pulse with $\chi=0.1$.
We first show the distribution functions $f_{nk}(t)$ for $n=0$, 1 and 2
versus $t$ and $k$ for a one-pulse excitation in Fig.\ 2
for $B=12$\ T. Here and hereafter, the additional transverse relaxation
time $T_2$ is taken as 300\ fs. The energy conserving
$\delta$-functions in the collision terms are replaced by
Gaussian functions with a width $\sigma=0.66$\ meV which is much smaller
than both $\omega_c$ and the inhomogeneous broadening. Note again
that $k$ is limited by Eq.\ (\ref{limitk}). One sees how the
carriers relax at later times in the lower $k$-states of the
Landau subbands. We plot the various
carrier densities $N_i(t) = \sum_k f_{ik}(t)$ against time $t$ for the
magnetic fields $B=12$\ T (Fig.\ 3(a)) and 18\ T (Fig.\ 3(b)).
For $B=12$\ T, one
sees pronounced relaxation oscillations of the populations of the
lowest subbands $N_0(t)$ and $N_1(t)$
in the first 400\ fs. The period of these relaxation oscillations
($\approx 60$\ fs) with successive overshoots in the population
of the Landau subband $n=0$ and $n=1$ is given by the characteristic
interband Coulomb scattering rate between these two lowest subbands.
The total carrier density $N_t=\sum_iN_i$ shows a slight Rabi flopping
overshoot. 
The incoherently summed polarizations $P(t)=\sum_{ik}|P_{ik}|$ for the two
fields are also plotted in the same figures. Strong quantum beats with
frequency $2\omega_c$ are revealed in the polarization for lower
magnetic fields up to 16\ T. This effect may be
measured by FWM. For $B>16$\ T, both the population oscillation and
the beating of the polarization amplitude die out (see Fig.\ 3(b)).
Moreover, one sees that in the latter case the carriers are
excited practically only in the lowest Landau subband.

In Fig.\ 4 we plot typical TR FWM signals versus
time $t$ for $B=12$\ T (Fig.\ 4(a)) and 18\ T (Fig.\ 4(b)) for
two pulses with the delay times $\tau=120$\ fs (solid
curve) and 240\ fs (dashed curve), respectively.
We find strong beating for lower magnetic fields, whereas
for higher fields the beating becomes weaker. Only very
small beats are found for $B=18$\ T in Fig.\ 4(b).
The beating totally disappears for $B=20$\ T. The
beating frequency for the TR FWM signal is exactly $2\omega_c$
for all magnetic fields where beating exists. From the
TI FWM signal as a function of the delay time $\tau$ one can obtain the
effective dephasing time.  Therefore
we plot the TI FWM signals in the same figures, however,
projected on the right and top scales.
Again one observes quantum beats in the TI FWM
signals for lower fields.  Similar beating
effect has been obtained based on a three-level model in
explaining heavy hole-light hole beats by Leo {\em et al.}.\cite{leo}
Moreover, the beats become weaker when the magnetic field increases
because then the second subband becomes less important.
For $B=16$\ T the beating has already disappeared in TI FWM which means that
the polarization components $P_{1k}$ are negligible in comparison to $P_{0k}$ .
This is consistent with Fig.\ 3(b), where the electron
(hole) population in the second Landau subband is extremely low and the
beating structure disappears in the incoherently summed polarization.
The beating frequency for the lower field is also $2\omega_c$. Note
again the different scales used for $t$ and $\tau$.
The dotted lines in both figures are the exponential fits to the
TI FWM signal. From the fit one gets the effective dephasing time
which is plotted as a function of the magnetic field $B$ in Fig.\ 5.
Interestingly, we find for
large magnetic fields that the dephasing
time decreases with increasing field for 10\ T$<B<16$\ T.
However, there is a transition at 16\ T. For $B>16$\ T the
dephasing time increases with increasing field, although much slower
than it decreased before.
\begin{figure}[hbt]
\psfig{figure=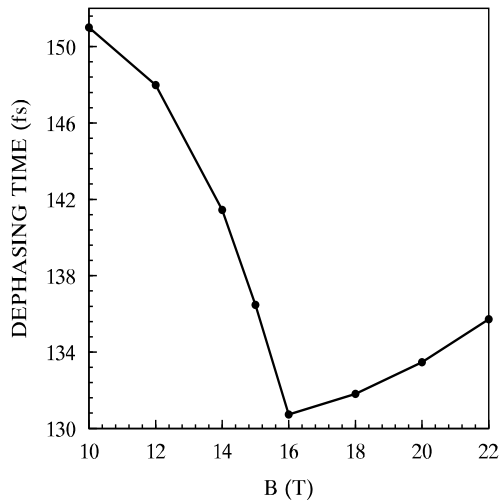,width=8.5cm,height=7.5cm,angle=0}
\caption{Dephasing time as a function of $B$.}
\end{figure}

For a fixed pulse,  several effects compete with each other
when the magnetic field increases. On one hand,
the number of Landau subbands which contribute to the Coulomb scattering
kinetics decreases. In particular the contributions to the dephasing from the
intra- and inter-subband scattering of the higher Landau subbands as well as
the inter-subband scattering between the higher and lower subbands decrease.
For large populations in one subband the Pauli blocking may further
reduce also the intra-band scattering rates. All these effects
increase the dephasing time. On the other hand, with increasing $B$
field the degeneracy of Landau subbands increases and the matrix elements of
the Coulomb scattering of Eq. (\ref{v}) become larger. How an increasing
degeneracy increases the scattering rates can be seen
from Eqs.\ (\ref{fscatt}) and (\ref{pscatt}) in the extreme limit where
the confinement is lifted. Then there is no $k$ and $k^\prime$
dependence in either distribution functions or polarization functions.
Therefore $\sum_{k^\prime}$ is replaced
by $Sm\omega_c/(2\pi)$ ($S$ standing for the 2D area)\cite{callaway}
which increases as $B$ increases. Such an ef-
\begin{figure}[bth]
\psfig{figure=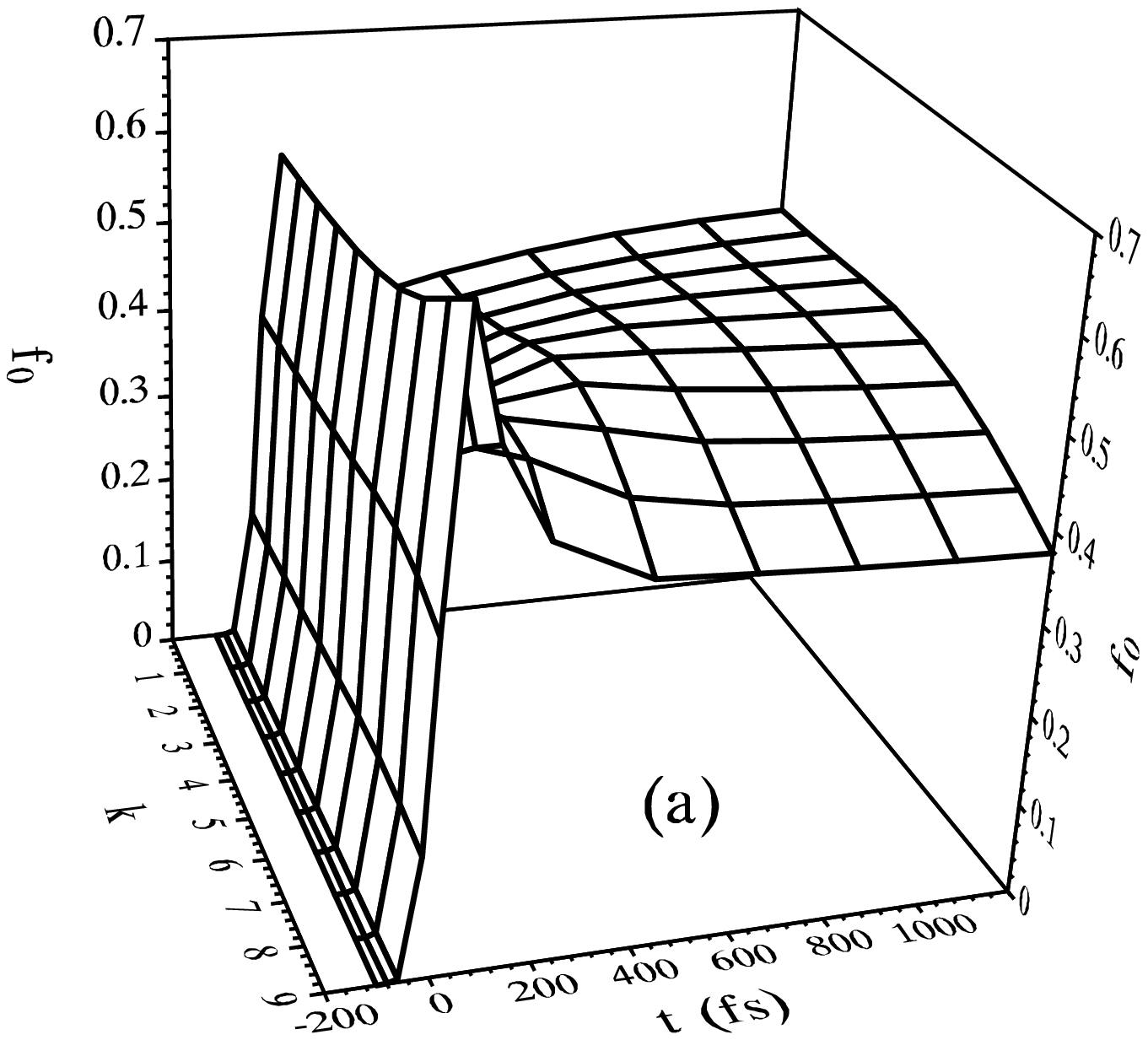,width=8.5cm,height=6.5cm,angle=0}
\psfig{figure=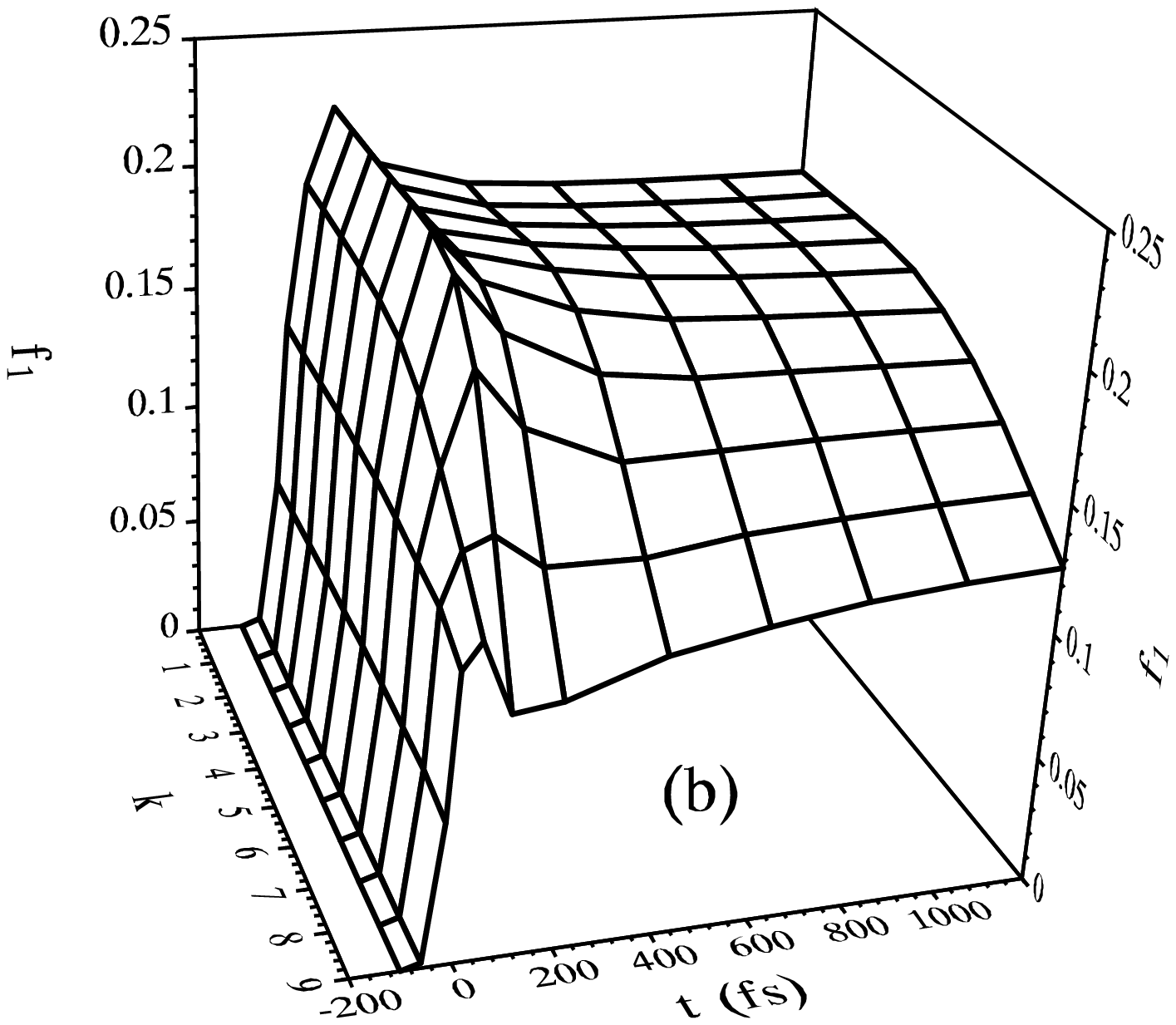,width=8.5cm,height=6.5cm,angle=0}
\psfig{figure=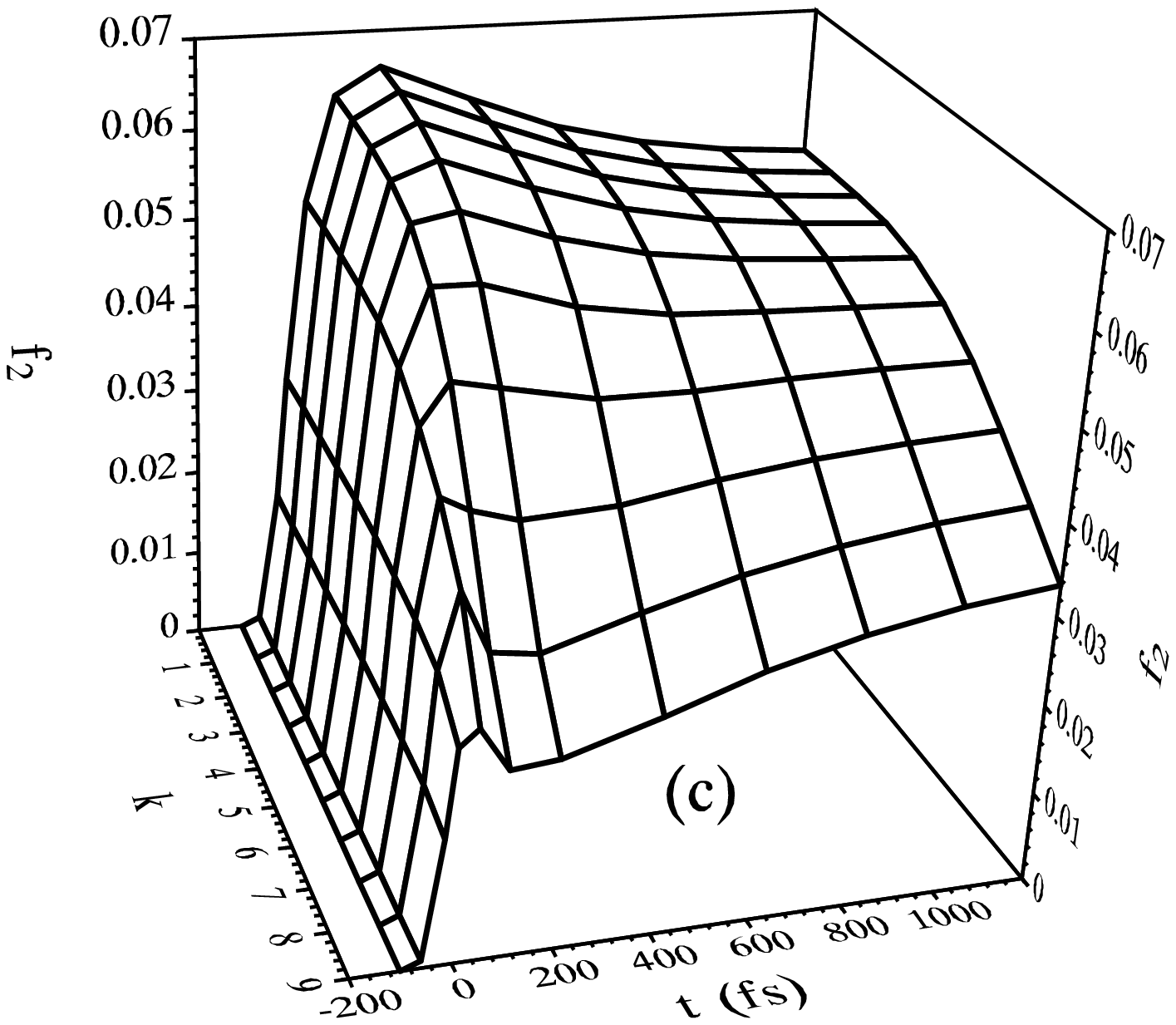,width=8.5cm,height=6.5cm,angle=0}
\caption{Distribution functions of three Landau subbands $f_0$, $f_1$
and $f_2$ versus $t$ and $k$ for one-pulse
excitation for $B=12$\ T. The units of $k$ in these
figures are one tenth of the maximum allowed value of $k$ defined in
Eq.\ (\ref{limitk}).}
\end{figure}
\noindent fect is also kept after
the weak broadening in $k$ space is included. Thus both the increased
degeneracy and the increased Coulomb matrix elements reduce the dephasing
time. Particularly, for $B \ge 16$T the above discussed increase in the
scattering rates is overcompensated by the loss of the contributions of the
inter-subband scattering processes and the reduction of the intra subband
scattering due to Pauli blocking, so that a net slow increase of the
dephasing time with increasing B field is obtained. For 10T $\le B\le$ 16T
the increasing degeneracy and the increasing Coulomb matrix elements dominate
over the reduction of inter-subband scattering so that the dephasing time
decreases with increasing B field.
\begin{figure}[htb]
\psfig{figure=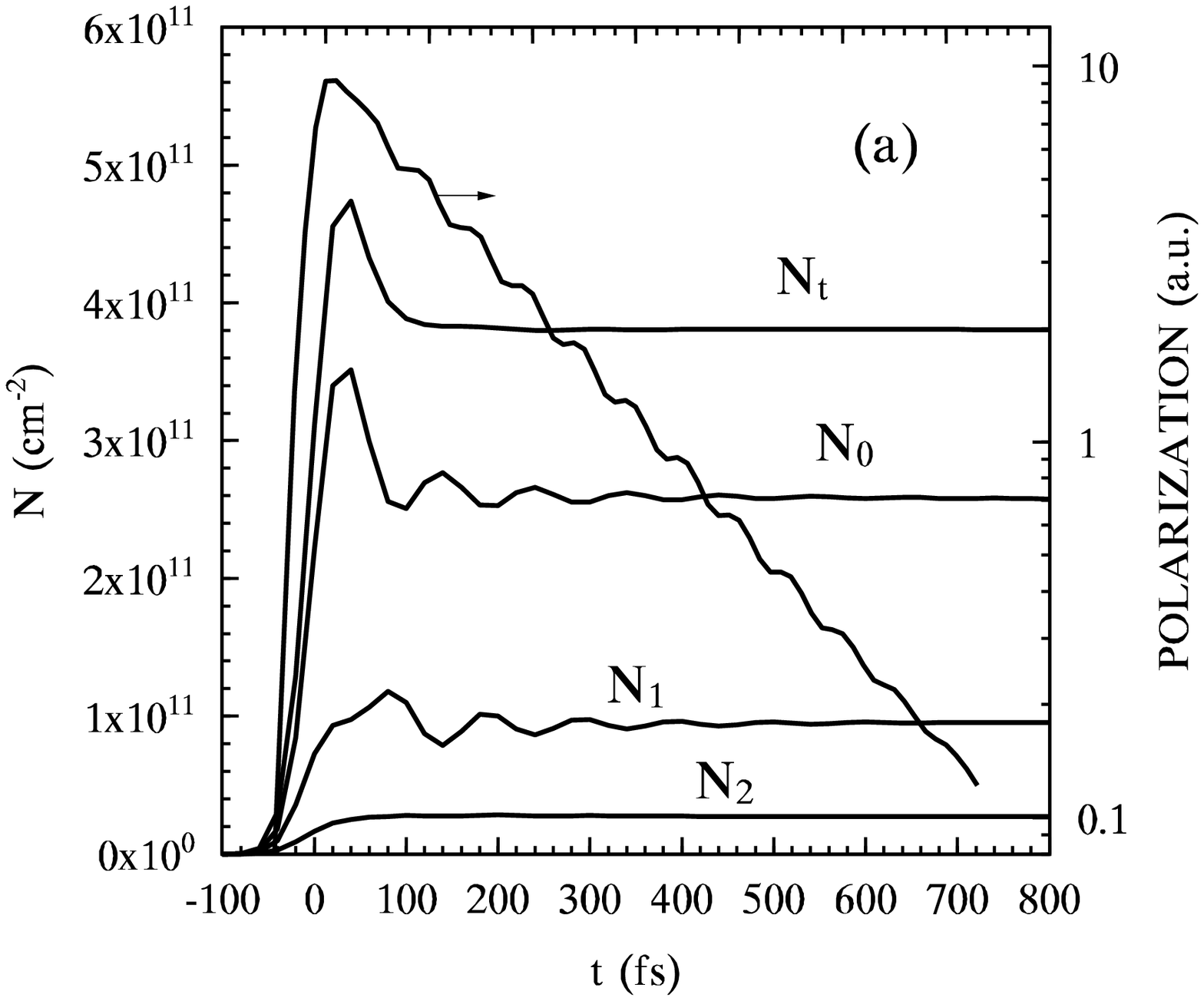,width=8.5cm,height=7.5cm,angle=0}
\psfig{figure=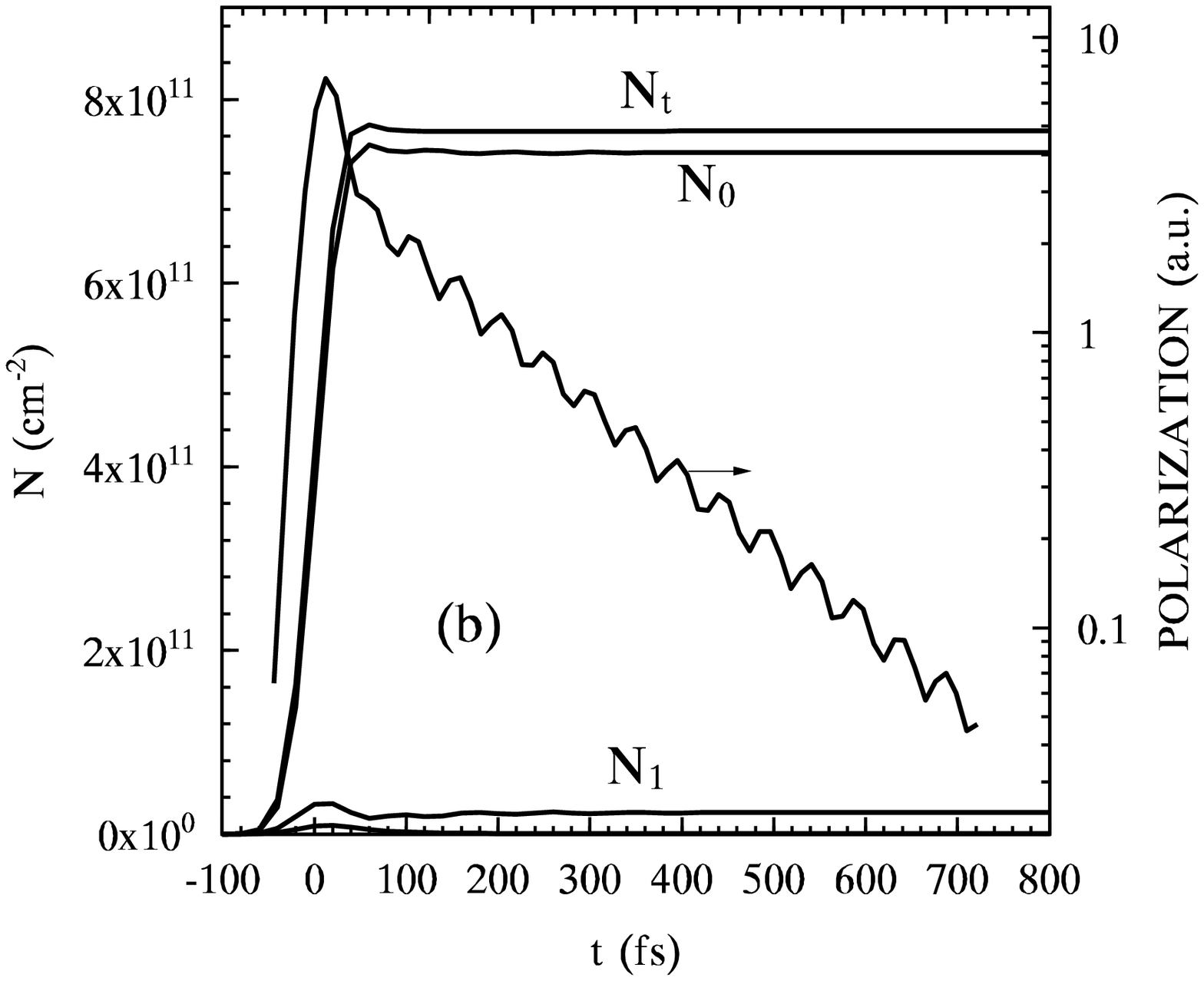,width=8.5cm,height=7.5cm,angle=0}
\caption{Electron density and incoherently summed polarization
are plotted against time $t$ for $B=12$\ T (Fig.\ 3(a)) and 18\ T
(Fig.\ 3(b)). $N_n$ ($n=0$, 1, and 2) is carrier density in
$n$-th Landau subband. $N_t=N_1+N_2+N_3$ is the total density.}
\end{figure}

\subsection{High-density case}

In order to obtain a dense magneto-plasma with
strong carrier interaction, we
choose $\chi=0.3$ corresponding roughly to a $\pi/3$ pulse,
which provides a very high excitation density. The detuning is
kept to be $\Delta_0=26.4$\ meV. Note that the local field
renormalization will actually turn the optical Bloch vector through
an angle considerably larger than $\pi/3$.
We show the distribution function $f_{nk}(t)$ with $n=0$, 1 and 2
versus $t$ and $k$ for one-pulse excitation in Fig.\ 6
for $B=12$\ T.
\begin{figure}[bth]
\psfig{figure=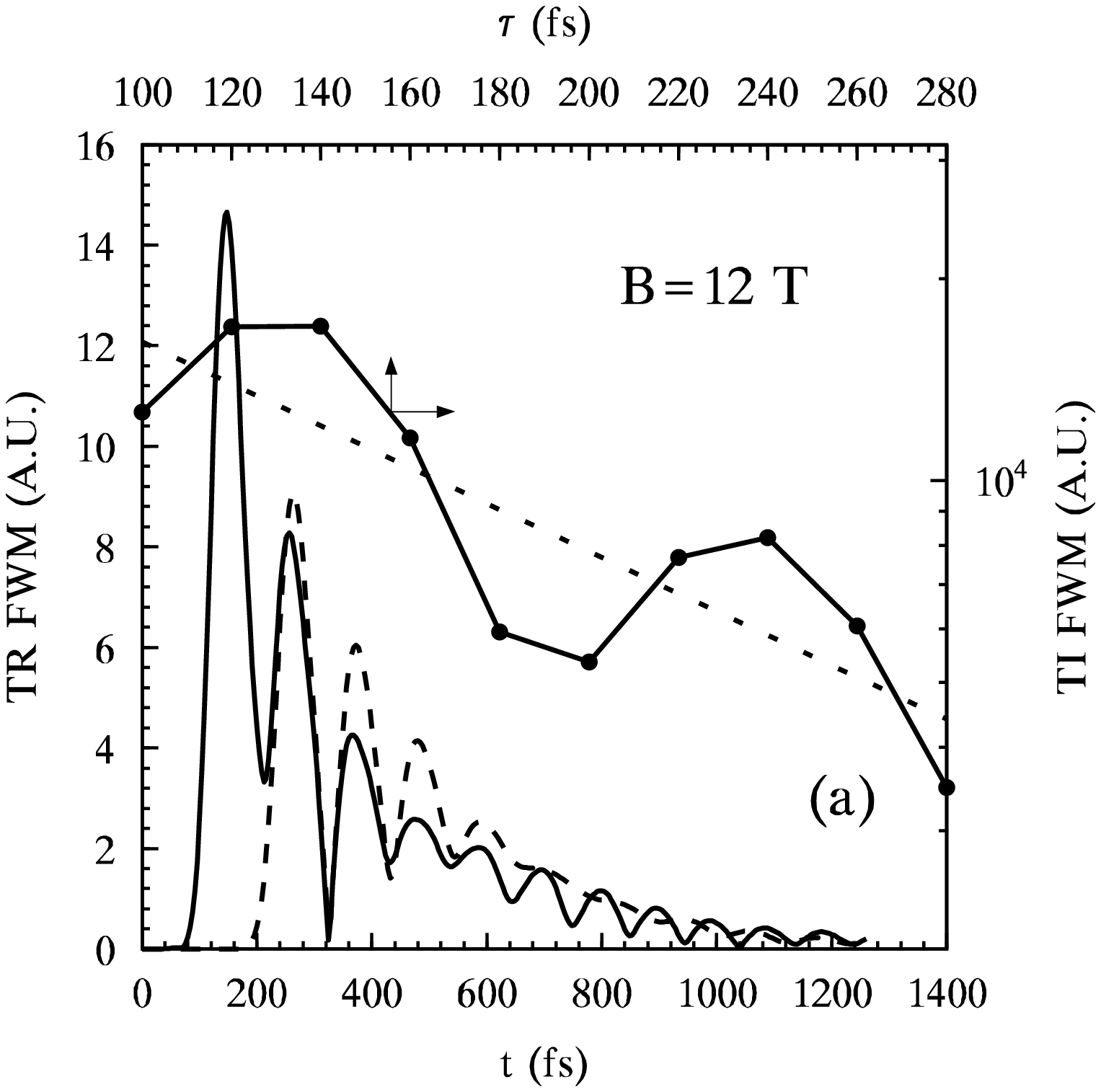,width=8.5cm,height=7.5cm,angle=0}
\psfig{figure=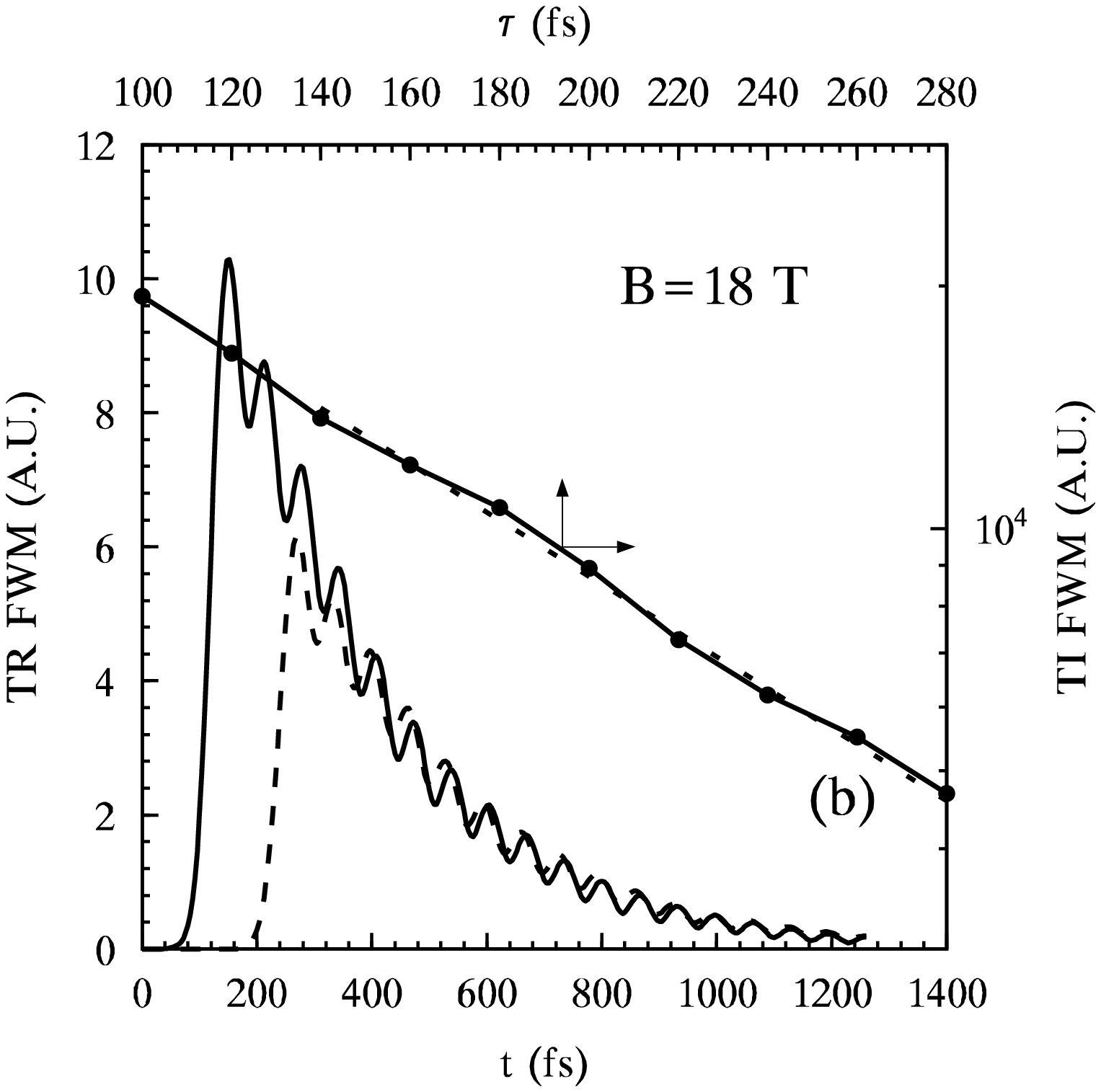,width=8.5cm,height=7.5cm,angle=0}
\caption{TR and TI FWM signals versus time $t$ and delay $\tau$
respectively for $B=12$\ T (Fig.\ 4(a)) and 18\ T (Fig.\ 4(b)).
For TR FWM, the solid curve is for $\tau=120$\ fs and dashed
curve for 240\ fs. The dotted line along TI FWM signal is
the exponential fit to the TI FWM signal.}
\end{figure}

 From the figure one can see that the carrier distribution
function for the lowest
Landau subband is close to 0.5. This number approaches 1 when
$B$ is around 18\ T. One sees a pronounced Rabi-flopping in the
two lowest Landau subbands. We further plot the
carrier density against time $t$ for two different magnetic
fields $B=12$\ T (Fig.\ 7(a)) and 18\ T (Fig.\ 7(b)). The incoherently
summed polarizations for the two fields are also plotted in the
same figures. Differing from the intermediate-density case,
strong quantum beats are revealed for
both field strengths, while the population relaxation oscillations
with a period of about 100\ fs
are again only well resolved for the low-magnetic field case ($B=12$\ T).
Again, the lowest Landau subband is always excited most strongly.

In Fig.\ 8 we plot again typical TR FWM signals against
time $t$ for $B=12$\ T (Fig.\ 4(a)) and 18\ T (Fig.\ 4(b)) for
two pulses with the delay times $\tau=120$\ fs (solid
curve) and 240\ fs (dashed curve), respectively. Similar to the
intermediate-density case, we find quantum beating for fields as
high as 20\ T with a frequency of exactly $2\omega_c$,
although for large fields the beating
become weaker. The TI FWM signals are plotted in the same figures as
functions of delay time $\tau$ (upper time scale).
Again one observes quantum beats in the TI FWM signals.
Interestingly, the beating frequency is close to that
in the incoherently summed polarization and is a little bit smaller than
$2\omega_c$ for higher fields. For example, when $B=18$\ T,
$\pi/\omega_c=66.5$\ fs
whereas from the beating in TI FWM, one finds a period
around 76\ fs. This difference may be understood in terms of many-body
effects, such as band gap renormalization, Pauli blocking  and excitonic
enhancement. We conclude that the quantum beats in the high-density case are
also caused by interferences between the optically induced polarizations 
$P_{n=0k}$ and $P_{n=1,k}$ of the two lowest subbands.

The effective dephasing is plotted as a function of the magnetic
field $B$ in Fig.\ 9. Once more we find a gradual overall
decrease of the dephasing time with increasing field in the
regime from 10 to 20\ T. This is in agreement with our
earlier results in the intermediate-density case for 10\ T$<B<16$\ T.
The dephasing is mainly controlled in the whole range by the kinetics
in the lowest two Landau subbands. Compared to the
intermediate-density regime, one observes a sharp dip around 15\ T in the
dephasing time.
By investigating the distribution function, we find the dip around 15 T
is mainly due to the function of the second pulse. When $B$ is
smaller than 15\ T, the second pulse further pumps electrons
from valence band to conduction band. Due to the density dependent red 
shift of the subbands and the Pauli blocking for the optical transition in 
the lowest subband, the second pulse increases the population of the 
second subband relative to that of the first one.
The ratio of the densities in
the lowest two subbands $N_0/N_1$ falls from around 5.7 after
the first pulse to 4 after the second pulse when $B$=14\ T,
and that number goes from 8 to 3.9 for $B=15$\ T. This means the role
of second subband becomes more important to the dephasing. This
gives rise to the sharp decrease of the dephasing time.
However for $B>15$\ T, the lowest Landau subband is
already highly populated and due to the
high magnetic field value, the second Landau subband is
already relatively far
away from the center of the pulse, so that
the second pulse depopulates the electrons from the first conduction
subband. Although this makes again the density of lowest subband 
closer to the second one, the ratio $N_0/N_1$ after the second pulse
goes from 3.9 at 15\ T to 6 at 16\ T. This clearly indicates  the
loss of the second subband to the dephasing processes and explains the fast
increase of the dephasing time from 15 to 16\ T. After 16\ T, although the
first pulse populates more carriers to the lowest subband
with increasing magnetic field due to the density-dependent 
band gap shift, the rate of change
becomes smaller as the lowest subband is nearly filled. However, 
the depopulation
due to the second  pulse becomes stronger with increasing magnetic fields.
This can be seen from the fact that $N_0/N_1$ after the second
pulse goes down from 6 for $B=15$\ T to finally 5 for $B=20$\ T. Our
results show that the decrease of Pauli blocking 
together with the increasing of intra-subband scattering matrix element
in the lowest subband dominate 
in the high magnetic field range the dephasing and explain the
resulting  decrease of dephasing time.
\begin{figure}[htb]
\psfig{figure=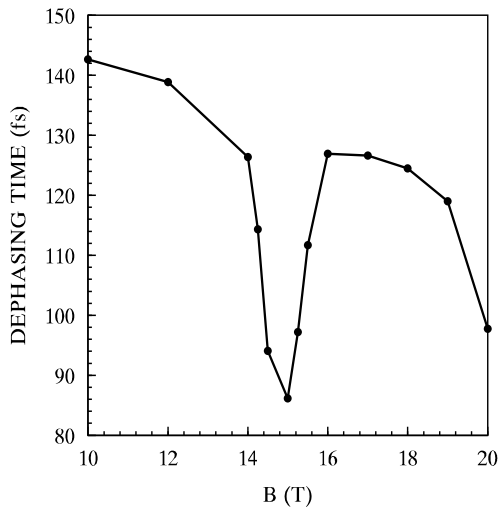,width=8.5cm,height=7.5cm,angle=0}
\caption{Dephasing time as a function of $B$.}
\end{figure}

\section{comparison with quantum kinetic calculation}

Because of the assumed pulse duration of 50\ fs is already
shorter than period of our quantum beats $T_c=2\pi/(2\omega_c)=119.7$\ fs
at $B=10$\ T, one may expect
that the memory effects of the quantum kinetic scattering integrals already
may play a role. Therefore we performed also quantum kinetic
calculation based on Eqs.\ (\ref{fscatt})
and (\ref{pscatt}). In this case, there is an undetermined
damping $\Gamma$ in the memory kernel. In principle this term should be
function of $k$, $q$, the magnetic field and the electron
distributions, and should be
determined self-consistently.\cite{hh} However, the
inclusion of such effects will complicate the quantum kinetics
considerably. Therefore, we took $\Gamma$ as an adjustable constant.
Such a simple approximation  causes the violation of energy
conservation in longtime limit and we find that the results,
particularly for the TI FWM signals and therefore the
dephasing are highly sensitive to the actual value of the damping constant.

In Fig.\ 10 we plot the TR FWM signal for pulses with $\chi=0.3$
and $\tau=240$\ fs for $B=18$\ T. The solid curve is based on
quantum kinetics with $\Gamma=1.32$\ meV. The dashed
curve is the prediction based on Boltzmann kinetics as shown in
Fig.\ 8(b). From Fig.\ 10 one can see that the two
theories give qualitatively
similar TR FWM signals, although the coherence of the polarization components
is clearly larger according to quantum kinetics in the first few hundred
femtoseconds because of the retarded onset of the dephasing processes in this
theory.
\begin{figure}[htb]
\psfig{figure=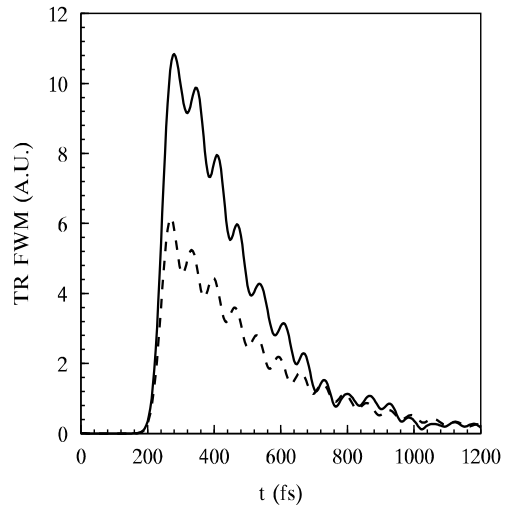,width=8.5cm,height=7.5cm,angle=0}
\caption{TR FWM for pulses with $\chi=0.3$ and delay $\tau=240$\ fs.
$B=18$\ T. Solid curve: quantum kinetic result; Dashed curve: Boltzmann
kinetics result.}
\end{figure}

However, we point out that the consistence in Fig.\ 10
depends on the choice of $\Gamma$ and we failed to get consistent
results for all the fields and delays with {\em one} constant
damping $\Gamma$. A more detailed treatment of the damping
of the memory kernel is needed for a consistent quantum kinetic theory.

\section{conclusion}

In conclusion, we have performed theoretical studies of
femtosecond kinetics of an optically excited 2D magneto-plasmas
for intermediate- and high-density excitations. Based on a
three-subband model in both the conduction and the valence band,
our study is restricted to high magnetic fields.
We calculated the intra- and inter-Landau-subband kinetics by
bare Coulomb potential scattering and found pronounced relaxation
oscillations to occur in the population of the two lowest Landau subbands
for lower magnetic fields. We calculated both TR
and TI FWM signals. Both signals
exhibit quantum beats with frequencies around $2\omega_c$.
Surprisingly, the resulting dephasing times are rather short and
are modulated in the strong magnetic field by about 30\%. Depending
on the detuning of the laser pulses, the pulse width, the excited densities
and the number of excited Landau subbands one gets decrease of the dephasing
time with increasing $B$ field because of the increase of the Coulomb matrix
elements and the degeneracy of the Landau subbands. In regions where the
loss of scattering channels exceeds these increasing effects, one can
also obtain a slight increase of the dephasing time with the magnetic field.
It is shown that the retarded onset of the dephasing and relaxation processes
in quantum kinetics result in a FWM signal  which is more coherent in the
first few hundred femtoseconds. However a consistent theory of the decay of
the memory kernel in a magneto-plasma is still missing. Finally, we
point out that more Landau subbands are needed to account for the
kinetics in lower magnetic fields. A corresponding extension of the theory is
still under investigation and the results will be published elsewhere.

\acknowledgements

We acknowledge financial support by the DFG within the
DFG-Schwerpunkt ``Quantenkoh\"arenz in Halbleiter''. MWW would like to
thank Prof. L. B\'anyai for valuable discussions during the initial stage
of this work which helped him to get into this filed smoothly.

\appendix

\section{}

We give the analytic expressions for all the Coulomb interaction
matrix elements Eq.\ (\ref{v}) up to the third Landau subbands.
We first show the approximation we take in our calculation through
$V_{00;00}$. After integrating out $x$ and $x^\prime$,
$V_{00;00}$ of Eq.\ (\ref{v}) can be written into
\begin{eqnarray}
V_{00;00}&=&\frac{e^2}{\epsilon_0} e^{-\frac{1}{2}\alpha^2\delta x_q^2
-\alpha^2\lambda^2(k-k^\prime)^2}\nonumber\\
&&\mbox{}\times\int_{-\infty}^\infty \frac{e^{-z^2}dz}
{\sqrt{(z-i\alpha\lambda(k-k^\prime))^2+q^2/(2\alpha^2)}}\;.
\label{v0000}
\end{eqnarray}
Due to the factor $e^{-\alpha^2\lambda^2(k-k^\prime)^2}$
$V_{00;00}$ decays quickly with increasing $|k-k^\prime|$. Therefore the
main contribution to the matrix element comes from small values of
$|k-k^\prime|$. We therefore neglect the term
$i\alpha\lambda(k-k^\prime)$ inside the integrand of Eq.\ (\ref{v0000}).
The integral remaining can be carried out analytically. Similar
approximation may be applied to other matrix elements and
therefore, Eq.\ (\ref{v}) may be written as
\begin{eqnarray}
&&V_{ni;jm}(q,k,k^\prime)\nonumber\\
&\simeq&\sum_{q_x}\frac{2\pi e^2}{\sqrt{q^2+q_x^2}}
\int d(x_1x_2)e^{-iq_x(x_1-x_2)-\alpha^2\lambda^2(k-k^\prime)^2}
\nonumber\\
&&\mbox{}\hspace{0.3cm}\times
\phi_n^\ast (x_1+\delta x_q)
\phi^\ast_i(x_2-\delta x_q)\phi_j (x_2)\phi_m (x_1)\;.
\end{eqnarray}
Defining two dimensionless variables $x=q^2/4m\Omega_x$ and
$y=2x\omega_c^2/\Omega_x^2$ as Ref.\ \onlinecite{bayer}, one can
express the matrix elements in terms of zeroth- ($K_0$) and first-order
($K_1$) modified Bessel functions. As there exist the following
symmetry relations for the matrix elements
\begin{eqnarray}
V_{ni;jm}&=&V_{jm;ni}
=(-1)^{n+m+i+j}V_{in;mj}\nonumber\\
&=&(-1)^{n+m+i+j}V_{mj;in}\;,
\label{symm}
\end{eqnarray}
one only needs to calculate 27 terms. The remaining
terms may be got from the symmetry relations. In the following we
give these required 27 terms:
\[V_{00;00}=\frac{e^2}{\epsilon_0}e^{x-y-\Xi}K_0(x)\;,\]
with $\Xi=\alpha^2\lambda^2(k-k^\prime)^2$.
\[V_{10;00}=-\frac{e^2}{\epsilon_0}\sqrt{y}e^{x-y-\Xi}K_0(x)\;.\]
It is noted here and hereafter the term $\sqrt{y}$ strictly should
be $\sqrt{y}\mbox{sgn}(q)$ with sgn standing for the sign function.
However, due to the fact the all the terms with the factor $\sqrt{y}$
do not appear in the Hartree-Fock terms, we therefore just
simplify it as $\sqrt{y}$.
\[V_{10;10}=\frac{e^2}{\epsilon_0}e^{x-y-\Xi}[(y-x)K_0(x)+xK_1(x)]\;,\]
\[V_{11;00}=\frac{e^2}{\epsilon_0}e^{x-y-\Xi}[-(x+y)K_0(x)+xK_1(x)]\;,\]
\[V_{10;01}=\frac{e^2}{\epsilon_0}e^{x-y-\Xi}[(1+x-y)K_0(x)
-xK_1(x)]\;,\]
\[V_{11;01}=\frac{e^2}{\epsilon_0}e^{x-y-\Xi}\sqrt{y}[(1+x+y)K_0(x)
-xK_1(x)]\;,\]
\begin{eqnarray*}
V_{11;11}&=&\frac{e^2}{\epsilon_0}e^{x-y-\Xi}\{[1+2x+2x^2+y(y-2x-2)]K_0(x)
\\
&&\mbox{}+x(2y-1-2x)K_1(x)\}\;,
\end{eqnarray*}
\[V_{20;00}=\frac{e^2}{\sqrt{2}\epsilon_0}e^{x-y-\Xi}[(x+y)K_0(x)-
xK_1(x)]\;,\]
\[V_{12;00}=-\frac{e^2}{\epsilon_0}e^{x-y-\Xi}\sqrt{y/2}[(3x+y)K_0(x)
-3xK_1(x)]\;,\]
\[V_{02;01}=\frac{e^2}{\epsilon_0}e^{x-y-\Xi}\sqrt{y/2}
[(y-x)K_0(x)+xK_1(x)]\;,\]
\end{multicols}
\widetext
\[V_{12;01}=\frac{e^2}{\sqrt{2}\epsilon_0}e^{x-y-\Xi}[(x+y+2x^2-
y^2)K_0(x)-2x^2K_1(x)]\;,\]
\[V_{02;02}=\frac{e^2}{\epsilon_0}e^{x-y-\Xi}[(\frac{1}{2}y^2-xy+x^2)K_0(x)+
(xy+\frac{1}{2}x-x^2)K_1(x)]\;,\]
\[V_{00;22}=\frac{e^2}{\epsilon_0}e^{x-y-\Xi}[(\frac{1}{2}y^2+3xy+x^2)
K_0(x)+(\frac{1}{2}x-x^2-3xy)K_1(x)]\;,\]
\[V_{20;21}=\frac{e^2}{\epsilon_0}e^{x-y-\Xi}\sqrt{y}[(\frac{1}{2}y^2
+x^2+x-y-xy)K_0(x)+(xy-\frac{1}{2}x-x^2)K_1(x)]\;,\]
\[V_{10;22}=-\frac{e^2}{\epsilon_0}e^{x-y-\Xi}\sqrt{y}[(\frac{1}{2}y^2
-3x-y+xy-3x^2)K_0(x)+(3x^2-xy+\frac{3}{2}x)K_1(x)]\;,\]
\[V_{22;20}=\frac{\sqrt{2}e^2}{\epsilon_0}e^{x-y-\Xi}[(\frac{1}{4}y^3
-\frac{1}{4}xy^2-\frac{1}{2}x^2y+x^3-y^2+\frac{7}{4}x^2+\frac{1}{2}x)K_0(x)
+(\frac{1}{4}xy^2+\frac{1}{2}x^2y-x^3-\frac{1}{4}xy-\frac{5}{4}x^2)K_1(x)]\;,
\]
\[V_{20;01}=\frac{e^2}{\epsilon_0}e^{x-y-\Xi}\sqrt{\frac{y}{2}}
[(y-x-2)K_0(x)+xK_1(x)]\;,\]
\[V_{12;10}=\frac{e^2}{\sqrt{2}\epsilon_0}e^{x-y-\Xi}[(y^2-2y-2x-2x^2)
K_0(x)+(x+2x^2)K_1(x)]\;,\]
\[V_{02;11}=-\frac{e^2}{\sqrt{2}\epsilon_0}e^{x-y-\Xi}[(y^2-2xy
+2x^2+2x-2y)K_0(x)+(2xy-2x^2-x)K_1(x)]\;,\]
\[V_{11;21}=\frac{e^2}{\epsilon_0}e^{x-y-\Xi}\sqrt{\frac{y}{2}}[(3y
-3x-2-y^2+2xy-2x^2)K_0(x)-(2xy-2x^2-2x)K_1(x)]\;,\]
\[V_{11;22}=-\frac{e^2}{\epsilon_0}e^{x-y-\Xi}\{[\frac{1}{2}(y-2)
(y^2-2y-2x-xy-2x^2)+\frac{3}{2}x^2+2x^3]K_0(x)+(\frac{1}{2}xy^2
+x^2y-\frac{1}{2}xy-x-2x^3-\frac{5}{2}x^2)K_1(x)\}\;,\]
\begin{eqnarray*}
V_{21;22}&=&\frac{e^2}{\epsilon_0}e^{x-y-\Xi}\sqrt{2y}[(\frac{1}{4}y^3-
\frac{3}{2}y^2+3xy+\frac{3}{2}x^2y-\frac{3}{4}xy^2+\frac{5}{2}y
-\frac{5}{2}x-1-\frac{5}{2}x^2-2x^3)K_0(x)\\
&&\mbox{}+[x^3+\frac{3}{4}y^2x
-\frac{3}{2}x^2y-\frac{9}{4}xy+\frac{3}{2}x+\frac{9}{4}x^2)K_1(x)]\;,
\end{eqnarray*}
\[V_{02;20}=\frac{e^2}{\epsilon_0}e^{x-y-\Xi}[(1+2x-2y-xy+x^2+
\frac{1}{2}y^2)K_0(x)+(xy-x^2-\frac{3}{2}x)K_1(x)]\;,\]
\[V_{12;20}=-\frac{e^2}{\epsilon_0}e^{x-y-\Xi}\sqrt(y)
[(1+2x-2y-xy+x^2+\frac{1}{2}y^2)K_0(x)+(xy-x^2-\frac{3}{2}x)K_1(x)]\;,\]
\begin{eqnarray*}
V_{12;21}&=&\frac{e^2}{\epsilon_0}e^{x-y-\Xi}[(1+3x-3y-5xy+
\frac{5}{2}y^2+\frac{9}{2}x^2-\frac{1}{2}y^3+2x^3-3x^2y+\frac{3}{2}xy^2)
K_0(x)\\
&&\mbox{}+(-\frac{3}{2}x+\frac{7}{2}xy-\frac{7}{2}x^2-2x^3-\frac{3}{2}
xy^2+3x^2y)K_1(x)]\;,\end{eqnarray*}
\begin{eqnarray*}
V_{21;21}&=&\frac{e^2}{\epsilon_0}e^{x-y-\Xi}[(\frac{1}{2}y^3-2y^2
+2y-\frac{3}{2}xy^2+4xy-2x+3x^2y-\frac{7}{2}x^2-2x^3)K_0(x)\\
&&\mbox{}+(\frac{3}{2}xy^2-\frac{5}{2}xy-3x^2y+x+2x^3+\frac{5}{2}x^2)
K_1(x)]\;,\end{eqnarray*}
\begin{eqnarray*}
V_{22;22}&=&\frac{e^2}{\epsilon_0}e^{x-y-\Xi}[(\frac{1}{4}y^4
-4x^3y+3x^2y^2+2x^4-xy^3-2y^3+6xy^2-11x^2y+7x^3+5y^2-10xy+
\frac{35}{4}x^2-4y\\
&&\mbox{}+4x+1)K_0(x)
+(xy^3-3x^2y^2+4x^3y-2x^4-\frac{9}{2}xy^2+9
x^2y-6x^3+6xy-6x^2-\frac{3}{2}x)K_1(x)]\;.\end{eqnarray*}

\begin{multicols}{2}
\narrowtext

\end{multicols}

\end{document}